\documentclass[12pt,letterpaper]{article}
\pdfoutput=1
\usepackage[utf8]{inputenc}
\usepackage{amsmath}
\usepackage{slashed}
\usepackage{amssymb}
\usepackage{amsthm}
\usepackage{xcolor}
\definecolor{nicered}{rgb}{0.7,0.1,0.1}
\definecolor{nicegreen}{rgb}{0.1,0.5,0.1}
\usepackage[colorlinks=true,citecolor=nicegreen,linkcolor=nicered]{hyperref}
\usepackage{graphicx}
\graphicspath{{figures/}}
\usepackage{siunitx} 
\usepackage{cancel}
\usepackage{cite}
\usepackage{soul}

\title{
The \textit{Fermi}-LAT gamma-ray excess at the Galactic Center in the singlet-doublet
fermion dark matter model
}

\author{
Shunsaku Horiuchi\footnote{\href{mailto:horiuchi@vt.edu}{horiuchi@vt.edu}}, Oscar Macias\footnote{\href{mailto:oscar.macias@vt.edu }{oscar.macias@vt.edu}}\\
\textit{\small Center for Neutrino Physics, Department of Physics, }\\
\textit{\small Virginia Tech, Blacksburg, VA 24061, USA}\\[4mm]
Diego Restrepo\footnote{\href{mailto:restrepo@udea.edu.co}{restrepo@udea.edu.co}},  
    Andrés Rivera\footnote{\href{mailto:afelipe.rivera@udea.edu.co}{afelipe.rivera@udea.edu.co}},
    Oscar Zapata\footnote{\href{mailto:oalberto.zapata@udea.edu.co}{oalberto.zapata@udea.edu.co}}\\
\textit{\small  Instituto de F\'{i}sica, Universidad de Antioquia,} \\
\textit{\small  Calle 70 No. 52-21, Medell\'{i}n, Colombia}\\[4mm]
Hamish Silverwood\footnote{\href{mailto:h.g.m.silverwood@uva.nl}{h.g.m.silverwood@uva.nl}}\\
\textit{\small GRAPPA, University of Amsterdam }\\
\textit{\small Science Park 904, 1098 XH Amsterdam, The Netherlands}
}
\date{\small February 15, 2016}
\begin{document}

\maketitle
\begin{abstract}
The singlet-doublet fermion dark matter model (SDFDM) provides a good DM candidate as well as the possibility of generating neutrino masses radiatively.
 The search and identification of DM requires the combined effort of both indirect and direct DM detection experiments in addition to the LHC.
 Remarkably, an excess of GeV gamma rays from the Galactic Center (GCE) has been measured with the \textit{Fermi}  Large Area Telescope (LAT) which appears to be robust with respect to changes in the diffuse galactic background modeling.
 Although several astrophysical explanations have been proposed, DM remains a simple and well motivated alternative.
 In this work, we examine the sensitivities of dark matter searches in the SDFDM scenario using \textit{Fermi}-LAT, CTA, IceCube/DeepCore, LUX, PICO and LHC with an emphasis on exploring the regions of the parameter space that can account for the GCE.
 We find that DM particles present in this model with masses close to $\sim 99$ GeV and $\sim (173-190)$ GeV annihilating predominantly into the $W^+W^-$ channel and $t\bar{t}$ channel respectively, provide an acceptable fit to the GCE while being consistent with different current experimental bounds. We also find that much of the obtained parameter space can be ruled out by future direct search experiments like LZ and XENON-1T, in case of null results by these detectors. Interestingly, we show that the most recent data by LUX is starting to probe the best fit region in the SDFDM model.

\end{abstract}

\section{Introduction}
It is well established that Dark Matter (DM) makes up about $25\%$ of the energy density of the Universe and is about five times more abundant than ordinary matter \cite{hinshaw2012}.
  However, its fundamental nature remains mysterious.
 No known particle has the properties needed to constitute the DM, whose identity thus begs for new physics beyond the Standard Model (SM).
 Unveiling which particle accounts for the majority of the matter in the universe is a key open question at the interface of particle physics and cosmology.

A promising candidate for DM particles are weakly interacting massive particles (WIMPs).
 These are generally assumed to be at equilibrium in the early Universe, but then freeze out due to the rapid expansion of the Universe.
 If the WIMP masses are in the GeV to TeV range, and the annihilation cross sections are of order the weak interaction scale, the relic DM density measured by experiments today arises naturally~\cite{Funk:review}.

WIMP particles appear effortlessly in many extensions of the SM that resolve outstanding theoretical and phenomenological problems which are not necessarily related to the DM puzzle.
 In some of these models, WIMPs can be produced in high energy colliders (collider DM searches), elastically scatter off nuclei (direct DM searches) or annihilate and produce observable particles in astrophysical environments (indirect DM searches).
 High-energy photons in the gamma-ray ($\gamma$-ray) frequency is the most notable search channel of the later category, as they can travel almost unperturbed from their sources to the detectors.
 The Large Area Telescope on board the \textit{Fermi} satellite (\textit{Fermi}-LAT)~\cite{Fermi} is the most sensitive $\gamma$-ray detector in the few GeVs energy range.

At the bottom of the gravitational well of the Milky Way Galaxy, the Galactic Center (GC) is expected to be the region displaying the brightest emission of DM annihilations in the $\gamma$-ray sky~\cite{Funk:review}.
 However, a multitude of non-thermal astrophysical sources present in that region complicate the identification of a tentative DM signal~\cite{Funk:review}. Observations of the inner few degrees around the GC with the \textit{Fermi}-LAT have revealed an excess of $\gamma$-rays ~\cite{Goodenough2009gk,Vitale:2009hr,Hooper:2010mq,hooper,AbazajianKaplinghat2012,AbazajianKaplinghat2013,GordonMacias2013}. 
The spectrum of the Galactic Center excess (GCE) peaks at about 1-3 GeV and its spatial morphology is spherically symmetric varying with radius $r$ around the GC as $r^{-2\gamma}$ with  $\gamma\sim 1.2$. 
This emission has been found to extend out in Galactic latitude ($b$) up to about $|b|\lesssim 20^\circ$ ~\cite{hooperslatyer2013,Daylan:2014,CaloreCholisWeniger2015,TheFermi-LAT:2015kwa} and its presence appears  to be robust with respect to systematic uncertainties~\cite{GordonMacias2013,MaciasGordon2014,Daylan:2014,Zho2015,CaloreCholisWeniger2015,PorterMurgia2015,TheFermi-LAT:2015kwa}. 

There is an ongoing and intense debate as to what the origin of this signal is. A tentative explanation is an unresolved population of $\sim 10^3$ millisecond pulsars (MSPs)~\cite{Abazajian:2010zy,AbazajianKaplinghat2012,Wharton2012,GordonMacias2013,GordonMacias2013erratum,Mirabal2013,MaciasGordon2014,YuanZhang2014,BrandtKocsis2015,Lacroix:2015wfx,O'Leary:2016osi} or young pulsars \cite{OLeary2015,O'Leary:2016osi}.
 Nevertheless, some studies~\cite{Hooper2013,CholisHooperLinden2015,PetrovicSerpicoZaharijas2015,Lee2015,BartelsKrishnamurthyWeniger2015,Linden2015} have pointed out about the difficulties of reconciling this hypothesis with the GCE extending out as far as $\sim 10^\circ$ from the GC. On the other hand, recent works claim that the GCE is not smooth~\cite{Lee:2015fea,Bartels:2015aea}, and if confirmed, this would lend support to the MSPs alternative.
 Another scenario put forward is a series of energetic cosmic-ray injections in the GC \cite{CarlsonProfumo,Petrovic2014}.
 However, if the injected particles are mainly protons, it has been shown~\cite{MacGorCroProf2015} that this scenario is incompatible with the spatial morphology of the GCE in the inner $\sim 2^\circ$ of the Galaxy.
 In case the burst events contain protons as well as leptons, Ref~\cite{Cholis2015} finds suitable models that appear fine-tuned.

Despite these astrophysical uncertainties, a DM interpretation of the GCE cannot be ruled out yet~\cite{Goodenough2009gk,Hooper:2010mq,hooper,AbazajianKaplinghat2012,AbazajianKaplinghat2013,GordonMacias2013,MaciasGordon2014,Abazajian2014,Daylan:2014,Lacroix:2015wfx}.
   In this context, the spatial morphology of the GCE can be accommodated with a Navarro-Frenk-White (NFW) profile with a mildly contracted cusp of $\gamma\sim 1.2$, the measured spectrum implies a WIMP mass in the GeV energy range and an interaction cross section that coincides with the thermal relic cross section. 

A recent study of the GCE~\cite{CaloreCholisWeniger2015} selected a target region ($|b|>2^\circ$) that excluded the core of the GC. Additionally, the systematic uncertainties in the Galactic diffuse emission were estimated in a manner that made the low and high energy tails of the spectrum more uncertain than in previous analyses~\cite{GordonMacias2013,MaciasGordon2014,Abazajian2014,Daylan:2014}, which focused on a smaller region containing the inner $\sim 2^\circ$ of the GC. Although it is possible that the greater degree of uncertainty in the tails found by~\cite{CaloreCholisWeniger2015} is due to an intricate overlap of the GCE with the Fermi Bubbles~\cite{suslatyerfinkbeiner2010,Fermi-LAT:AnnaFranckowiak}, it is interesting that this uncertainty also allows much more freedom for DM models fitting the GCE~\cite{Logan:2010nw,Buckley:2010ve,Zhu:2011dz,Marshall:2011mm,Boucenna:2011hy,Buckley:2011mm,Anchordoqui:2013pta,Buckley:2013sca,Hagiwara:2013qya,Okada:2013bna,Huang:2013apa,Modak:2013jya,Boehm:2014hva,Alves:2014yha,Berlin:2014tja,Agrawal:2014una,Izaguirre:2014vva,Cerdeno:2014cda,Ipek:2014gua,Boehm:2014bia,Ko:2014gha,Abdullah:2014lla,Ghosh:2014pwa,Martin:2014sxa,Basak:2014sza,Berlin:2014pya,Cline:2014dwa,Han:2014nba,Detmold:2014qqa,Wang:2014elb,Chang:2014lxa,Arina:2014yna,Cheung:2014lqa,McDermott:2014rqa,Huang:2014cla,Balazs:2014jla,Ko:2014loa,Okada:2014usa,Ghorbani:2014qpa,Banik:2014eda,Borah:2014ska,Cahill-Rowley:2014ora,Guo:2014gra,Freytsis:2014sua,Heikinheimo:2014xza,Arcadi:2014lta,Richard:2014vfa,Cao:2014efa,Bell:2014xta,Hardy:2014dea,Fortes:2015qka,Cerdeno:2015ega,Cao:2015loa,Ghorbani:2015baa,Kim:2016Yeong,Caron:2015wda,Bertoneetal,Freeseetal}.

Significant effort has been made in exploring the properties of DM models that can explain the GCE while being consistent with other indirect, direct and collider constraints~\cite{Logan:2010nw,Buckley:2010ve,Zhu:2011dz,Marshall:2011mm,Boucenna:2011hy,Buckley:2011mm,Anchordoqui:2013pta,Buckley:2013sca,Hagiwara:2013qya,Okada:2013bna,Huang:2013apa,Modak:2013jya,Boehm:2014hva,Alves:2014yha,Berlin:2014tja,Agrawal:2014una,Izaguirre:2014vva,Cerdeno:2014cda,Ipek:2014gua,Boehm:2014bia,Ko:2014gha,Abdullah:2014lla,Ghosh:2014pwa,Martin:2014sxa,Basak:2014sza,Berlin:2014pya,Cline:2014dwa,Han:2014nba,Detmold:2014qqa,Wang:2014elb,Chang:2014lxa,Arina:2014yna,Cheung:2014lqa,McDermott:2014rqa,Huang:2014cla,Balazs:2014jla,Ko:2014loa,Okada:2014usa,Ghorbani:2014qpa,Banik:2014eda,Borah:2014ska,Cahill-Rowley:2014ora,Guo:2014gra,Freytsis:2014sua,Heikinheimo:2014xza,Arcadi:2014lta,Richard:2014vfa,Cao:2014efa,Bell:2014xta,Hardy:2014dea,Fortes:2015qka,Cerdeno:2015ega,Cao:2015loa,Ghorbani:2015baa,Kim:2016Yeong,Caron:2015wda,Bertoneetal,Freeseetal}.  
 Of great interest are the properties of minimal supersymmetric extensions of the SM (MSSM)~\cite{Cheung:2014lqa,Cahill-Rowley:2014ora,Cao:2014efa,Cerdeno:2015ega,Caron:2015wda,Bertoneetal,Freeseetal} that can fit the GCE.
 When these extensions are studied in light of the GCE extracted from the region $|b|>2^\circ$ of the GC, the required neutralino annihilation rates to mainly the $W^+W^-$ and $\bar{t}t$ channels are found to comply with the LEP or LHC bounds on sfermion masses.

Here, we do not restrict ourselves to supersymmetric models. Instead, we take the approach of studying a simplified DM model in which the DM candidate is a mixture, generated by the interaction with the Higgs boson, of a SM fermion singlet and the neutral components of an electroweak doublet vector-like fermion~\cite{ArkaniHamed:2005yv,Mahbubani:2005pt,D'Eramo:2007ga,Enberg:2007rp}. This model, also known as the singlet$-$doublet fermion DM (SDFDM) model, is one of the simplest UV realizations of the fermion Higgs portal ~\cite{Patt:2006fw} with the SM Higgs boson as the mediator between the visible and dark sectors. In fact, the dark sector of the SDFDM model (along with the stabilizing discrete symmetry) is part of the minimal setup expected when the SM is extended by new physics which is to some extent related to lepton and baryon number conservation~\cite{Arbelaez:2015ila,Arkani-Hamed:2015vfh}. While being free of many theoretical biases, this  model allows us to extract maximal phenomenological information from a framework that is a good representation of the WIMP paradigm \cite{ArkaniHamed:2005yv,Mahbubani:2005pt,D'Eramo:2007ga,Enberg:2007rp,Cohen:2011ec,Cheung:2013dua,Abe:2014gua,Calibbi:2015nha,Freitas:2015hsa,Abdallah:2015ter}\footnote{ If scalar singlets are added to its particle content, neutrino masses can also be radiatively generated in this generic class of models~\cite{Restrepo:2015ura}.}.     
Accordingly, the SDFDM model is set to become one of the models to be implemented in future searches for DM particles at the LHC \cite{Abdallah:2015ter} and a future 100 TeV hadron collider \cite{Gori:2014oua,Arkani-Hamed:2015vfh}.

In this article we examine the coverage of WIMP parameter space in the SDFDM model by using mainly indirect and direct DM search techniques in light of the recent detection of the GCE. We show the set of parameters in the SDFDM model that are compatible with the GCE while being consistent with current experimental bounds. Following the same methods explained in Ref.~\cite{Silverwood:2014yza} we compute the expected limits in the annihilation cross-section by the Cherenkov Telescope Array (CTA) and find that observations toward the GC by this instrument will not be able to confirm this model as an explanation of the GCE. However, we find that the viable models can be ruled out by future
direct search experiments such as LZ and XENON-1T, in the case of null results by these detectors. Interestingly, we show that the most recent data by LUX is starting to probe the best fit region in the SDFDM model. The rest of the paper is organized as follows: In section~\ref{sec:model}, we describe the SDFDM model and the dark matter production mechanisms. We provide details on the usage of the GCE data in Sec.~\ref{sec:gammarays_from_the_GC}, and our main results and conclusions are presented in Sec.~\ref{sec:Parameter-scan} and Sec.~\ref{sec:conclusions}, respectively.

\section{The SDFDM model}
\label{sec:model}
The particle content of the model consists of one singlet
Weyl fermion $N$ of hypercharge $Y=0$ and two $SU(2)_L$-doublets of
Weyl fermions $\Psi$ , $\Psi^c$ with hypercharges $Y=\mp 1/2$. 
These are odd under one imposed $Z_2$ symmetry, while the SM particles are even under the same discrete group. 
The most general $Z_2$-invariant Lagrangian contains the following  mass terms and Yukawa interactions
\begin{align}
\label{eq:lt13a}
 \mathcal{L}\supset & M_D \Psi\Psi^c-\tfrac{1}{2}M_N NN - y_1 H \Psi N - y_2 \widetilde{H}\Psi^c N+\text{H.c.},
\end{align}
where the new $SU(2)_L$-doublets are written in terms of the left-handed Weyl fermions $\Psi=(\psi^0,\,\psi^-)^{\operatorname{T}}$ and $\Psi^c=(-(\psi^-)^c,\,(\psi^0)^c)^{\operatorname{T}}$~\cite{Cohen:2011ec}, and the SM Higgs doublet is given by 
$H=(0,\,(h+v)/\sqrt{2})^{\operatorname{T}}$ with $\widetilde{H}=i\sigma_2H^*$ and $v=246$ GeV.

The $Z_2$-odd spectrum is composed by a
charged fermion $\chi^{\pm}$ with a tree
level mass $m_{\chi^\pm}=M_D$, and three Majorana fermions which arise from
the mixture between the neutral parts of the $SU(2)_L$ doublets and
the singlet fermion. 
Defining the fermion basis as
$\Xi=\left(N,\psi^0,(\psi^0)^c \right)^T$,
the neutral fermion mass matrix reads
\begin{align}
\label{eq:Mchi}
  \mathbf{M}_{\Psi}=\begin{pmatrix}
 M_N                 &-m_{\lambda}\cos\beta&m_{\lambda}\sin\beta\\
-m_{\lambda}\cos\beta &  0                  & -M_D\\
m_{\lambda}\sin\beta&  -M_D                &  0  \\
\end{pmatrix},
\end{align}
where $m_{\lambda}=\lambda v/\sqrt{2}$, $\lambda=\sqrt{y_1^2+y_2^2}$ and $\tan\beta=y_2/y_1$. 
In what follows, we assume CP invariance, which allows us to set $\tan\beta$ as a real parameter and $M_D,M_N$ and $\lambda$ to be positive. 
Moreover, we consider only $|\tan\beta|\geq1$ since the physics for $|\tan\beta|\leq1$ is equivalent. 
Importantly, the SDFDM model considered in this study acts as a limit of the minimal supersymmetric standard model when the winos are decoupled from the spectrum and 
$\lambda=g'/\sqrt{2}$. 

The Majorana fermion mass eigenstates $
\mathbf{X}=(X_1,X_2,X_3)^T$ are obtained through the rotation
matrix $\mathbf{U}$ as $\mathbf{X}=\mathbf{U}\mathbf{M}_{\Psi} $, such
that
\begin{align}
\label{eq:chidiag}
 \mathbf{U}\mathbf{M}_{\Psi}\mathbf{U}^{\operatorname{T}}=\mathbf{M}^\text{diag}_{\Psi},
\end{align}
where
$\textbf{M}^{\text{diag}}_{\Psi}=\operatorname{Diag}(m_1,m_2,m_3)$  (no mass ordering is implied) and $\mathbf{U}$ is a real mixing matrix. Here, the DM candidate is the lightest mass eigenstate $X_i$.
In order to compute the corresponding $m_i$ terms, we used the characteristic equation as given by
\begin{align}\label{eq:masses}
(M_N-m_i)(m_i^2-M_D^2)+m_\lambda^2(M_D\sin2\beta+m_i)=0.
\end{align}

At tree level, the interaction between the DM and the SM sector is mediated by the $W$, $Z$ and $H$ gauge bosons.
 In terms of the Majorana and Dirac spinors $\chi_i^0$ , $\chi^{\pm}$  \footnote{The corresponding  spinors are given by $\chi_i^0=(X_{i\alpha}  ,X_i^{\dagger\dot{\alpha}})^{T}$ and $\chi^+=(X^+_{\alpha} ,{X^-}^{\dagger\dot{\alpha}})^T$.}, the interaction terms can be written as
\begin{align}
\mathcal{L}\supset -& c_{h\chi_i\chi_j}h\bar{\chi}_i^0\chi_j^0
-c_{Z\chi_i\chi_j}Z_{\mu}\bar{\chi}_i^0\gamma^{\mu}\gamma^{5}\chi_j^0-\dfrac{g}{\sqrt{2}}(U_{i3}W_{\mu}^-\bar{\chi}_i^0\gamma^{\mu}P_L\chi^+-U_{i2}W_{\mu}^-\bar{\chi}_i^0\gamma^{\mu}P_R\chi^+ + \text{H.c.}),
\end{align}
where 
$c_{Z\chi_i\chi_j}=\frac{g}{4\cos\theta_W}(U_{i2}U_{j2}-U_{i3}U_{j3})$ and $c_{h\chi_i\chi_j}=\frac{1}{\sqrt{2}}(y_1U_{i2}U_{j1}+y_2U_{i3}U_{j1})$.  
As is usually done, we denote the lightest stable particle in our model by $\chi^0$, whose couplings are readily acquired from the latest set of equations. Explicitly, these are
\begin{align}
c_{Z\chi^0\chi^0}&=-\frac{m_Z\lambda^2v(m_{\chi^0}^{2}-M_D^2)\cos2\beta}{2(m_{\chi^0}^{2}-M_D^2)^2+\lambda^2v^2\left(2\sin2\beta m_{\chi^0} M_D+m_{\chi^0}^{2}+M_D^2\right)},\label{eq:cZXX}\\
c_{h\chi^0\chi^0}&=-\frac{(M_D\sin 2\beta+m_{\chi^0})\lambda^2v}{M_D^2+\lambda^2v^2/2+2M_N\,m_{\chi^0}-3m_{\chi^0}^{2}}\label{eq:cHXX}.
\end{align}
In our model, DM particles ($\chi^0$) can self-annihilate into $\bar{f}f$, $ZZ$, $W^+W^-$ and $hh$ final states through  $s$-channel Higgs and $Z$ boson exchange and into $ZZ$, $W^+W^-$ states via $t$-channel $\chi_i^0$ and $\chi^{\pm}$ exchange. Annihilations into a mixture of weak gauge bosons $Zh$ are also possible through the exchange of a $\chi_i\neq\chi^0$  in the $t$-channel or a $Z$ in the $s$-channel.  We remark in passing that gamma-ray lines $\gamma\gamma$ and $\gamma Z$ an also be produced at one-loop level. 

Of particular importance for indirect detection studies in this framework is the fact that since DM annihilations into fermion pairs mediated by the Higgs are $p$-wave suppressed (there is no $s$-wave amplitude), the annihilations produced through $Z$ exchange are dominant. We note that the later is also helicity suppressed, this implies that the main annihilation channel is the $t\bar{t}$ ($b\bar{b}$) for a dark matter mass above (below) the top mass, with $\langle\sigma v \rangle\lesssim 10^{-27}$ cm$^3$ s$^{-1}$ for $m_{\chi^0}<m_W$ \cite{Calibbi:2015nha}. In the case scenario of DM particles going into gauge bosons, we find that only those processes in the $t$-channel are relevant to our analysis as they do not suffer velocity suppression. Such a non-velocity suppression is also present in $s$ and $t$ channels for the annihilation into $Zh$. 
In contrast, we get that processes in which DM self-annihilates into a couple of Higgs bosons are velocity suppressed. 
At higher order in scattering theory the loop suppression leads to small values of the corresponding thermal cross sections \cite{Calibbi:2015nha}. One of the prime motivations of the present study is to explore the viable regions of the parameter space where the velocity averaged cross-section $\langle\sigma v\rangle$ can exhibit values comparable to those predicted by the WIMP paradigm. It is in this sense that we will not consider DM annihilations into $\gamma\gamma,\gamma Z,hh$ and $ b\bar{b}$ in the discussion that follows.

Regarding direct detection, the Higgs ($Z$) exchange leads to spin independent  (spin dependent) DM nucleon scattering. From Eq. (\ref{eq:cZXX}) we get that the spin dependent (SD) cross section vanishes for $\cos2\beta=0$ or $|m_{\chi^0}|=M_D$, implying for both cases that $\tan\beta=\pm1$. In the same vein, from Eq.~(\ref{eq:cHXX}) the spin independent (SI) cross section vanishes (i.e. a blind spot as discussed by Ref.~\cite{Cheung:2013dua}) for $\sin2\beta=-m_{\chi^0}/M_D$, which leads to $m_{\chi^0}=M_N, M_D$, via Eq.~(\ref{eq:masses}). Note that $\sigma_{SI}=0$ if $\tan\beta<0$ and that only if $M_N>M_D$ both $\sigma_{SI}$ and $\sigma_{SD}$ can be zero simultaneously. 

\section{Gamma-rays from the Galactic Center }
\label{sec:gammarays_from_the_GC}
The Galactic $\gamma$-ray intensity $\Phi(E_{\gamma},b,l)$ produced in self-annihilations of DM particles, where $b$ and $l$ are the Galactic latitude and longitude respectively, can be obtained from the following relation~\cite{Baltz, Bergstrom,Rott}
\begin{equation}
\Phi(E_{\gamma},b,l)= \frac{1}{2}\frac{\left\langle\sigma v \right\rangle}{4\pi m_{\chi^0}}\sum_{f} \frac{dN_{f}}{dE_{\gamma}}B_f \times J(b,l),
\label{Phi}
\end{equation}
which is the product of a term that depends solely on the inherent properties of the DM particle and an astrophysical factor $J(b,l)$ accounting for the amount of DM in the line of sight.
 The former is given in terms of the velocity-averaged annihilation cross-section $\left\langle \sigma v \right\rangle$, the differential $\gamma$-ray multiplicity per annihilation $dN_{f}/dE_{\gamma}$, the DM mass $m_{\chi^0}$ and the branching ratio $B_f$ where $f$ denotes the final state particles resulting from the annihilation. 
The astrophysical factor can be drawn as ~\cite{Bergstrom, Rott}
\begin{equation}
J(b,l)=\int^{\infty}_0 ds\; \rho\left(\sqrt{R^2_{\odot}-2sR_{\odot}\cos(b)\cos(l)+s^2}\right)^2,
\label{J}
\end{equation}
where the DM density-square is integrated along the line-of-sight $s$ and $R_{\odot}=8.25$ kpc is the distance from the solar system to the GC.

The DM halo density $\rho(r)$ is determined by N-body cosmological simulations, with recent studies preferring a generalized NFW profile~\cite{navarrofrenkwhite1997} of the form
\begin{equation}
\rho(r)=\frac{\rho_s}{\left(\frac{r}{r_s}\right)^{\gamma}\left[1+\left(\frac{r}{r_s}\right)^{\alpha}\right]^{(\beta-\gamma)/\alpha}},
\label{nfw}
\end{equation}
where we adopt the scale radius $r_s=23.1$ kpc and the parameters $\alpha=1$, $\beta=3$ as default choices.
 Recent analyses of the GCE~\cite{GordonMacias2013,Daylan:2014,CaloreCholisWeniger2015} find a best fit profile inner slope $\gamma \simeq 1.2$, corresponding to a mildly contracted DM halo.
 We normalized the density profile by fixing the local dark matter $\rho(R_\odot=8.25\; \mbox{kpc})=0.36$ GeV cm$^{-3}$.
 This was done by maximizing the likelihood of microlensing and dynamical data for the chosen profile slope (see Fig.5 of Ref.~\cite{ioccopatobertone2011}). 

The $\gamma$-ray spectra ($dN_{f}/dE_{\gamma}$) resulting from $\chi^0$ annihilations was generated with the software package \textsc{PPPC4DMID} \cite{Cirelli_cookbook}.
 We noticed that for some channels, the interpolation functions provided by this useful tool are incomplete close to the rest mass thresholds.
 In such cases, we instead generated the spectra with the Monte Carlo event generator \textsc{PYTHIA 8.1} \cite{Sjostrand:2007gs} making sure that these were in agreement with the ones in \textsc{PPPC4DMID} for higher mass ranges.
 
Because of the quadratic dependence of Eq.~\ref{Phi} on the dark matter density, the GC is expected to be the brightest DM source in the $\gamma$-ray sky. However this region also harbours many $\gamma$-ray compact objects and the Galaxy's most intense diffuse $\gamma$-ray emission produced by the interaction of cosmic rays with interstellar material. The impact of these uncertainties in the interpretation of the GCE is currently not very well understood and is the subject of many recent studies.

There are also large uncertainties associated with the predicted signal from DM self-annihilations in the GC. The DM distribution in the innermost region of our Galaxy is poorly constrained by numerical DM-only simulations and kinematic measurements of Milky Way constituents. In principle, ordinary matter is expected to affect the inner dark matter profile obtained from simulations at a certain level. The DM density could be either flattened by star burst activity that ejects baryonic material from the inner region or steepened through adiabatic contraction. Indeed, depending on the assumed DM distribution, different estimates of the expected $\gamma$-ray emission can differ by a factor of up to $\sim 50$ (see Ref.~\cite{Funk:review,Catena:2009mf}).

Dwarf spheroidal galaxies (dSph) of the Milky Way are generally thought to be much simpler targets for indirect DM detection. Although their  $J(b,l)$ factor is orders of magnitude lower than that of the GC, they contain a much cleaner $\gamma$-ray background. Reference~\cite{Ackermann:2015zua} shows that the null detection of $\gamma$-ray emission from such objects impose strong constraints on the properties of DM models. In the next sections, we will discuss the effects of these limits on the DM interpretation of the GCE. 

Here we entertain the possibility that the SDFDM model can account for the GCE while being consistent with a variety of experimental limits on DM. This is accomplished by following closely the procedure developed in Ref.~\cite{CaloreCholisWeniger2015} and expanded upon in  Ref.~\cite{Caron:2015wda,Bertoneetal}. In summary, the $\gamma$-ray fluxes obtained from our model scans are compared to the GCE data made available in Ref.~\cite{CaloreCholisWeniger2015}. In that work, the systematic and statistical uncertainties in the Galactic diffuse emission model were provided in the form of a covariance matrix $\Sigma_{ij}$, which we use here to the full extent (we refer the reader to the aforementioned article for details on the statistical formalism and the implementation of the $\chi^2$ function). As was done in Refs.~\cite{Caron:2015wda,Bertoneetal}, we modified the covariance matrix to also account for theoretical uncertainties in the $\gamma$-ray spectra generation. Namely, we rewrite $\Sigma_{ij}$ as 
\begin{equation}
\Sigma_{ij} \rightarrow \Sigma_{ij} + \delta_{ij} d^2_{i} \sigma^2_s,
\end{equation}
where $\delta_{ij}$ is the Kronecker delta, $d_i$ are the measured photon fluxes and $\sigma_s=10$\% is the adopted theoretical uncertainty~\cite{Caron:2015wda,Bertoneetal}.

For each of the SDFDM models, we calculate the corresponding $\chi^2$ (or $p$-value) and make sure that these are consistent with the null \textit{Fermi}-LAT detection of $\gamma$-rays in dSphs. As recommended in the 3FGL catalog article~\cite{Acero:2015hja}, a given source spectral model is rejected when its associated $p$-value is less than $10^{-3}$. This is the same as to say that for $24-4$ degrees of freedom ($d.o.f$), model points having a $\chi^2>45.37$ are considered bad fits to the GCE. In all relevant figures, we incorporate the 95\% upper limits on the value of $\left\langle\sigma v \right\rangle$ as extracted from Ref.~\cite{Ackermann:2015zua}.

\section{Numerical analysis}
\label{sec:Parameter-scan}

Having identified the main annihilation channels and established the procedure to calculate the $\gamma$-ray fluxes, we move to explore the regions of the parameter space that can account for the {\it Fermi} GeV excess. Namely, in this section we determine the regions that are compatible with current constraints coming from colliders, electroweak phase transition (EWPT), indirect and direct DM searches, and then assess them in light of the quality of the fit to the GCE.     

\subsection{Scan and constraints}

To this end, we scan the parameter space of our model by considering the following ranges for the model parameters:    
\begin{align}\label{eq:scan}
100<M_D/\text{GeV}<1000,&\hspace{1cm}10 <M_N/\text{GeV}<1000,\nonumber\\
10^{-4}<\lambda<10,&\hspace{1cm} 1\leq\left|\tan\beta\right|<60.
\end{align}

Essentially, we throw darts into this large space, generating several million random model points, and for each generated point we compute the DM relic density and the direct and indirect DM observables using \textsc{micrOMEGAs 4.1.8}~\cite{Belanger:2014vza} through \textsc{Feynrules 2.3}~\cite{Christensen:2008py}. Each individual model is then subjected to a large set of dark matter, precision measurement and collider constraints. In particular, we assume that the DM  relic density saturates the Planck measurement $\Omega h^2=(0.1199 \pm 0.0027)$~\cite{Ade:2013zuv} at the $3\,\sigma$ level as we are interested in considering in considering the case where this model accounts for the majority of DM. The model points are also required to be compatible with \textit{Fermi}-LAT constraints coming from dwarf spheroidal galaxies~\cite{Ackermann:2015zua}, as well as LUX~\cite{Akerib:2013tjd}, IceCube~\cite{2013PhRvL.110m1302A}, PICO-2L~\cite{Amole:2016pye} and PICO-60~\cite{Amole:2015pla}  limits for spin independent and spin dependent detection studies. 
Since the SDFM model presents new contributions to the EW precision observables (EWPO) \cite{D'Eramo:2007ga}, we impose the condition that $\Delta T < 0.2$ given that the contribution to $S$ is always negligible \cite{Calibbi:2015nha}.  Finally, the limit obtained from searches of charged vector-like particles by LEP~\cite{ALEPH:2005ab} has been taken into account by imposing the condition $M_D > 100$~GeV in Eq. (\ref{eq:scan}).

\begin{figure}[h]
\begin{center}
\includegraphics[scale=0.35]{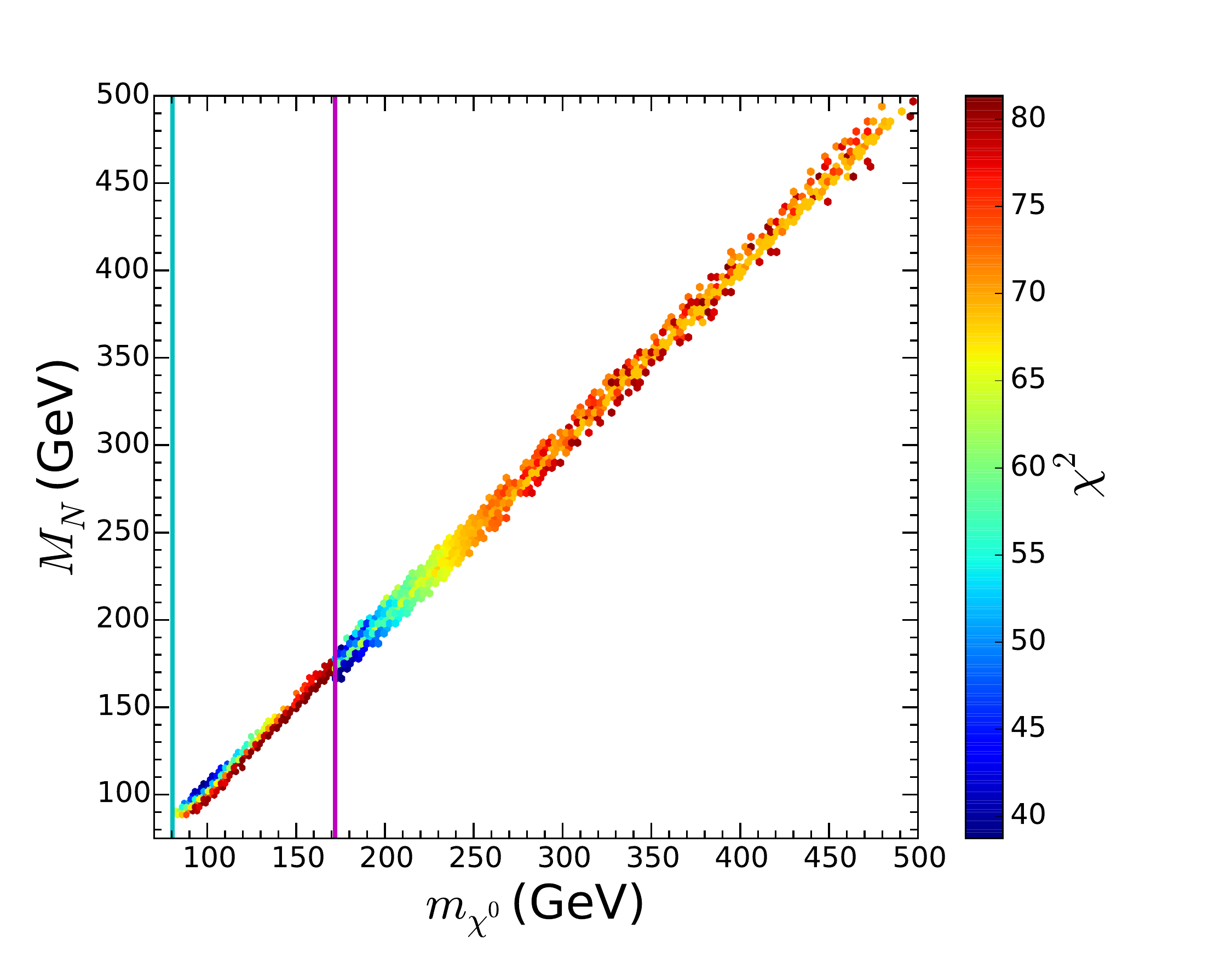}  \includegraphics[scale=0.35]{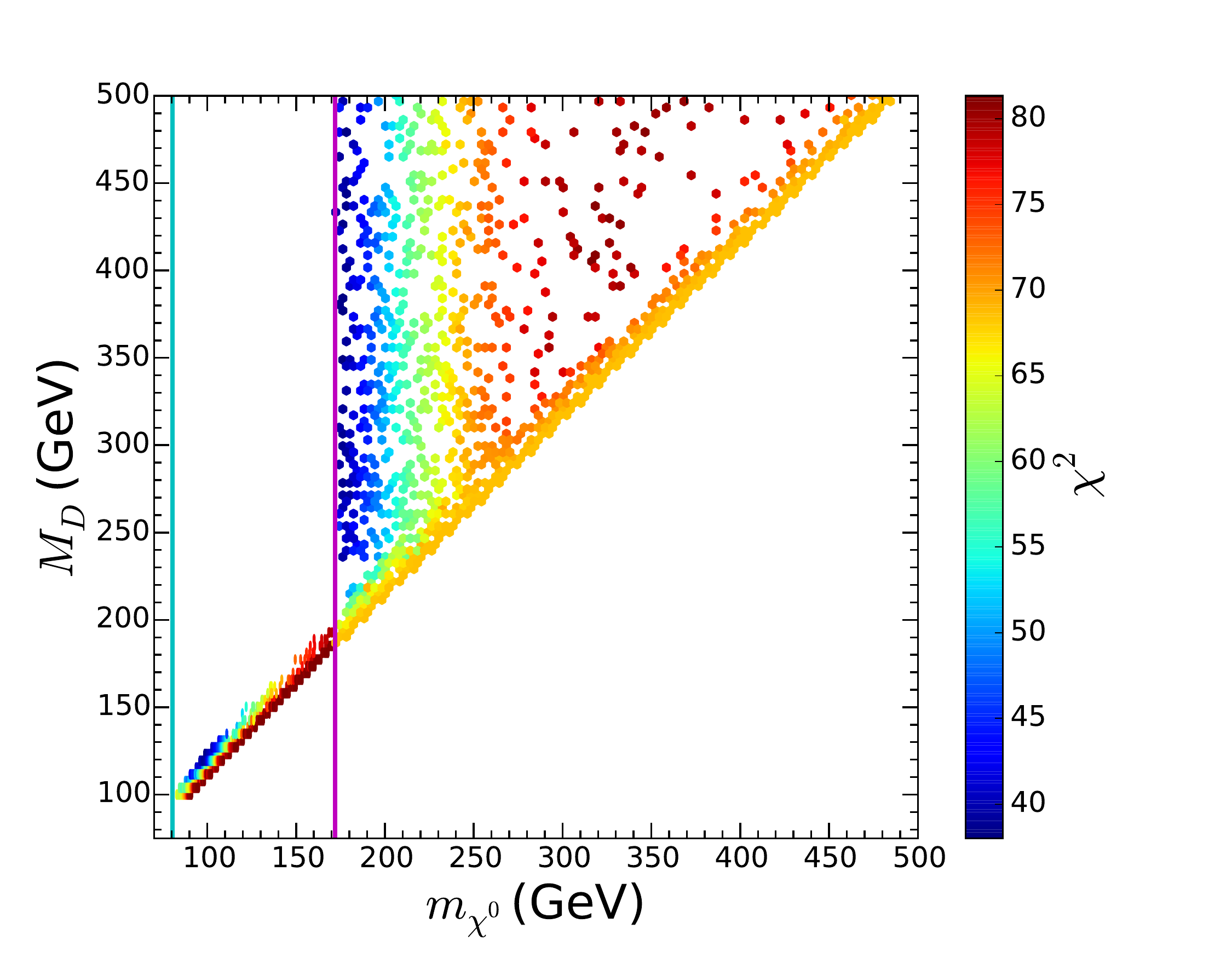} 
\includegraphics[scale=0.35]{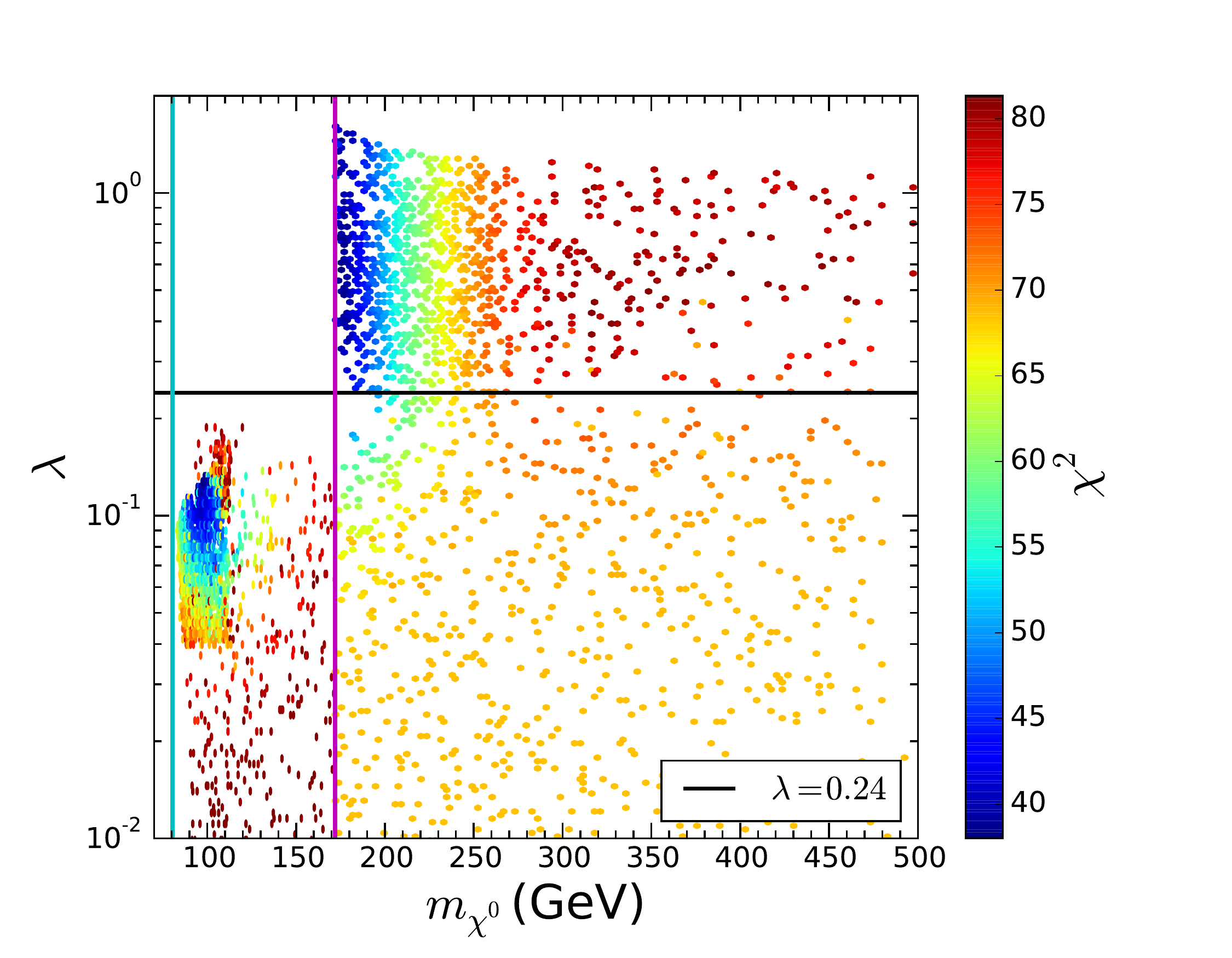}  \includegraphics[scale=0.35]{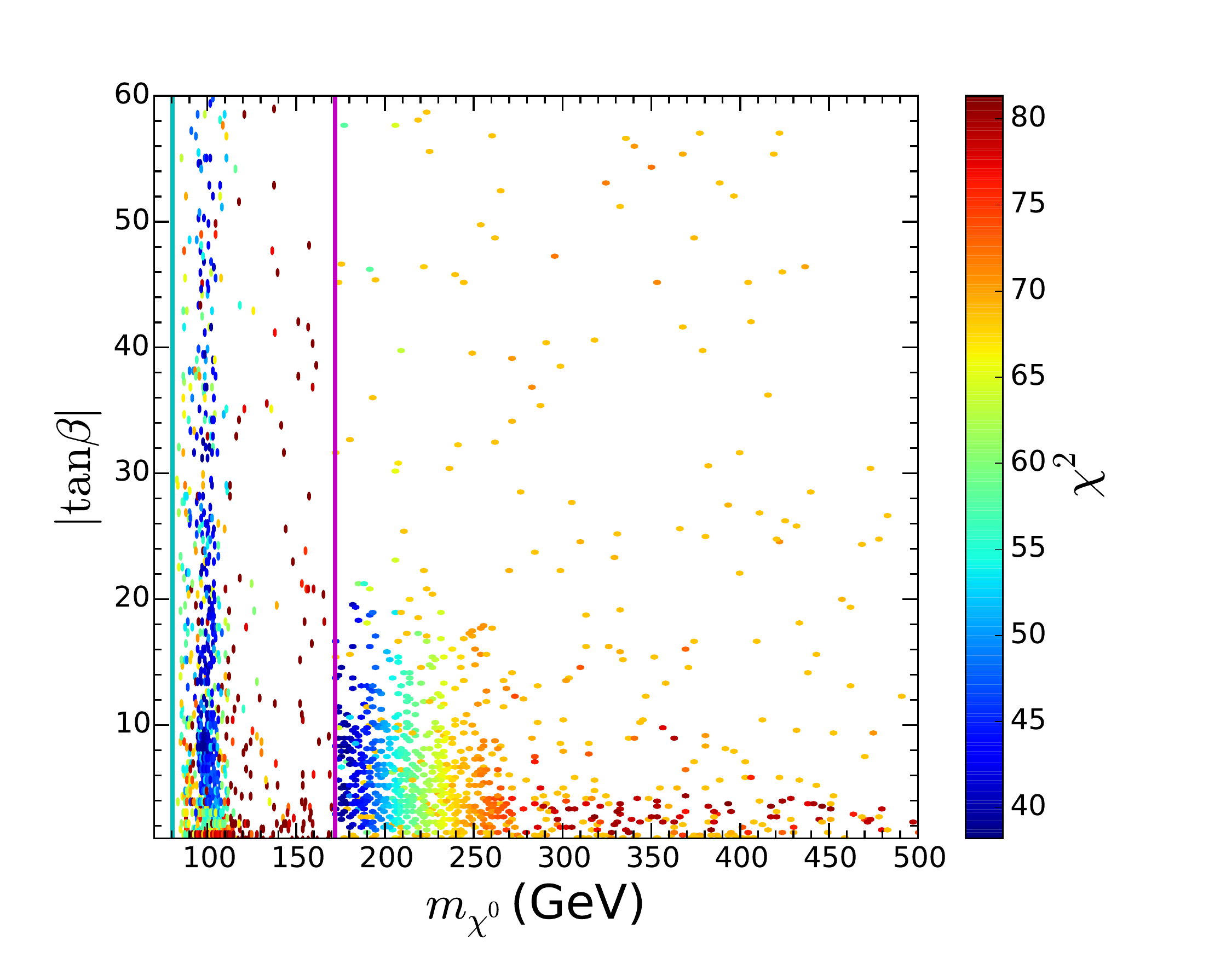}
\caption{Two-dimensional projection of the $\chi^2$ values of our fit, showing each one of the four free parameters in the SDFDM model ($M_N$, $M_D$, $\lambda$ and $|\tan \beta|$) versus the dark matter mass ($m_{\chi^0}$). 
In the bottom left panel the black line represents the supersymmetry value $\lambda\sim 0.24$, while the cyan and magenta vertical lines in all panels represent the $W$ boson mass and the top quark mass, respectively. Model points able to fit the GCE are those having a $\chi^2<45.37$ for $24-4$ d.o.f.}
\label{fig:parameters}
\end{center}
\end{figure}

\subsection{Results}
\label{subsec:Results}
Fig.~\ref{fig:parameters} displays the viable models in the planes ($M_N$, $m_{\chi^0}$),  ($M_D$, $m_{\chi^0}$),  ($\lambda$, $m_{\chi^0}$) and  ($|\tan\beta|$, $m_{\chi^0}$), along with the corresponding $\chi^2$ values obtained from a fit to the GCE. 
Since the fit tends to be worse for large values of $m_{\chi^0}$, we only considered DM masses  below 500 GeV. 
Furthermore, as it was discussed in Sec.~\ref{sec:model}, we only studied models with $m_{\chi^0}$ above the $W$ gauge boson mass. 
It is convenient to split the results of our scan into two different regions (DM mass ranges): one in which $m_{\chi^0}$ is below the top mass (Region I) and a second one in which $m_{\chi^0}$ is larger than the mass of the top quark (Region II).  

The viable models belonging to Region I are characterized for having  $M_N\approx M_D\approx m_{\chi^0}$, that is, the DM particle is a mixture of singlet and doublet states (well-tempered DM \cite{ArkaniHamed:2006mb,Cheung:2013dua}). 
The non-observation of direct detection signals constrains the Yukawa coupling to small values ($y<0.2$). We note that this limit excludes the MSSM value $\lambda \sim 0.24$. However, $|\tan\beta|$ is not constrained to a specific value or range. 
Regarding Region II, our analysis shows that $M_N\approx m_{\chi^0}$ while $M_D\gtrsim m_{\chi^0}$. For $y\lesssim 0.3$ the DM particle should be again well tempered ($M_D\approx M_N$) whereas for larger values of $y$ we have that $M_D$ is larger than $M_N$.      
In this case the upper bound $y\lesssim 5$ comes from the Planck measurement of the DM relic density.

The viable solutions to the GCE found in Region I feature the following parameters: $M_N\sim 105$ GeV, $M_D\sim 120$ GeV, $\lambda\sim 0.12$ and $|\tan\beta|\sim 9$ which generates a DM mass of $\sim 99$ GeV with a $\chi^2$ value of $45.3$. For these parameters the dark matter annihilates mostly into $W^+W^-$. 
While for the Region II we found that the viable solutions correspond to the sample:
\begin{align}
166  < & \, M_N/\text{GeV} < \, 197 , \nonumber \\
236  < & \, M_D/\text{GeV} < \, 988 , \nonumber \\ 
0.25  < & \, \lambda < \, 1.60 ,\nonumber \\
1.87 <  &\, \tan\beta < \, 19.6,
\end{align}
which leads to a DM mass in the range $(173-190)$ GeV with $\langle\sigma v\rangle_{t\bar{t}}/\langle\sigma v\rangle\geq 0.9$ and $\langle\sigma v\rangle_{WW}/\langle\sigma v\rangle\leq 0.1$. The fact that $\chi^0\chi^0\to t\bar{t}$ dominates, via $s$-channel exchange of a $Z$, is reflected in the required values for $y$, because it controls the coupling $c_{Z\chi^0\chi^0}$ whenever $|\tan\beta\neq1|$. 
Note also that, since $\tan\beta>0$ and $\tan\beta\neq1$, the SI and SD cross sections respectively can not be zero (no blind spot occurs). This means that the hypothesis of the SDFDM model being an explanation of the GCE can be probed in future experiments (see next section). Concerning the best $\chi^2$ obtained, we have obtained the value $38.0$ which is represented by white star in Fig.~\ref{fig:sigmav} and Fig.~\ref{fig:sigma-SI-SD}. 

Overall, the two sets of models capable of explaining GCE have DM particles $\chi^0$ with masses around $99$ GeV and $173-190$ GeV annihilating into $W^+W^-$ and $t\bar{t}$, respectively\footnote{The fact that the DM should annihilate into $W^+W^-$ and $t\bar{t}$ in order to explain the GCE is in accordance with what was stated in Ref.~\cite{Agrawal:2014oha}.}. As explained above, all of our solutions saturate the thermal relic density, making them also consistent with cosmological constraints on dark matter. 

\subsection{Probing the viable solutions with future observations}
\begin{figure}[t]
\begin{center}
\includegraphics[scale=0.5]{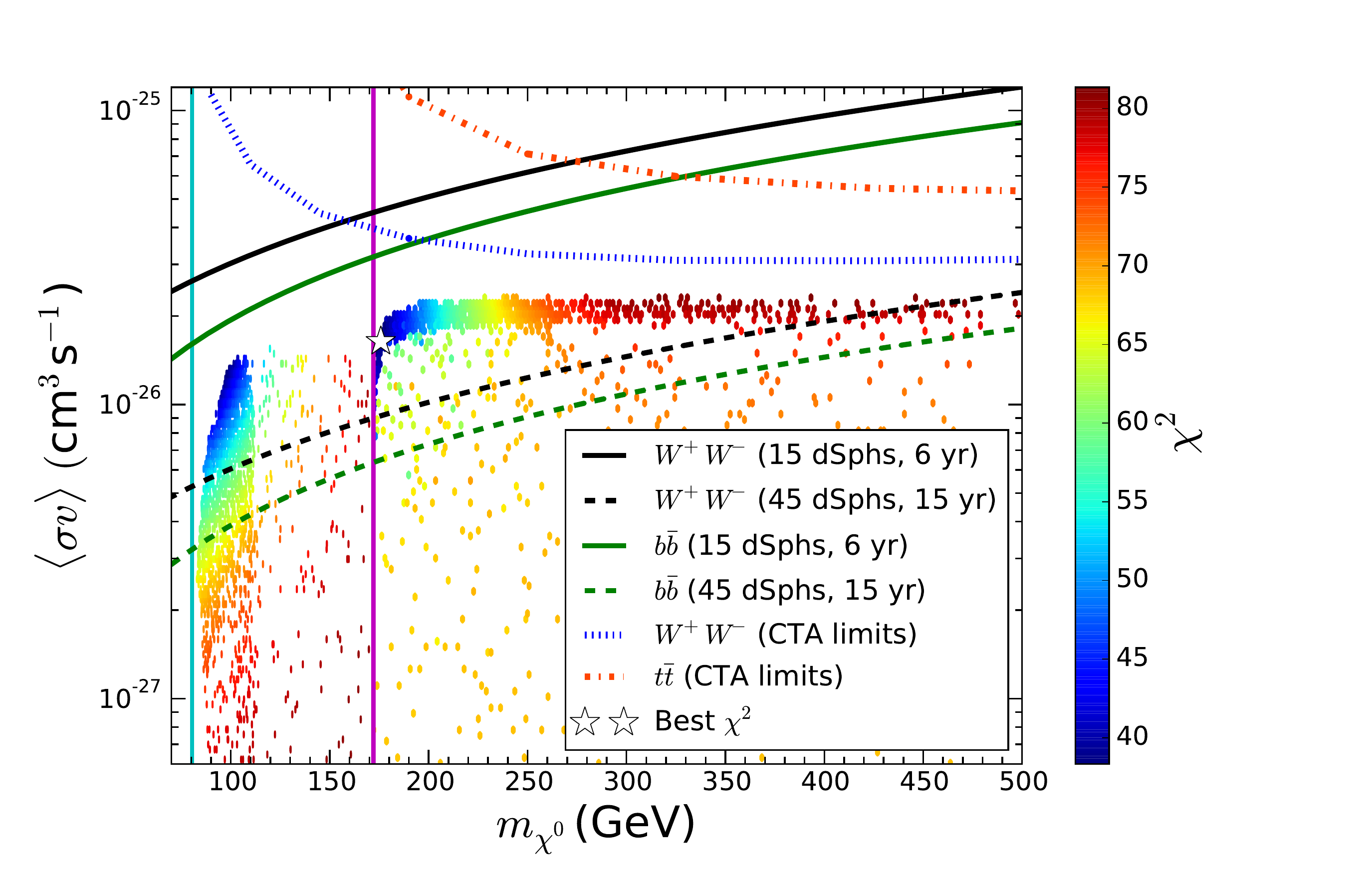} 
\caption{The present velocity averaged annihilation cross-section as a function of the dark matter mass in comparison to current indirect detection limits in different channels. 
The 95\% C.L gamma-ray upper limits from dSphs are extracted from Ref.~\cite{Ackermann:2015zua}. The CTA limits correspond to future 100 hr of $\gamma$-rays observations of the GC and assume a generalized NFW profile with an inner slope of $\gamma=1.2$.
The star is the best-fitting model obtained from our scan. 
Vertical lines and color code are the same as in Fig.~\ref{fig:parameters}.}
\label{fig:sigmav}
\end{center}
\end{figure}
The velocity averaged annihilation cross-section as a function of the dark matter mass in
comparison to current indirect detection limits in different channels along with the $\chi^2$ values found in a fit to the GCE are shown in Fig.~\ref{fig:sigmav}. 
Note that current upper limits from dSphs~\cite{Ackermann:2015zua} do not presently constrain any of the viable points. This is a consequence of the imposed requirement that models must comply with the observed DM relic density. Once this condition is applied, it generally restricts the parameter space of the SDFDM model to have a $\langle\sigma v\rangle$ less than $\sim 2\times 10^{-26}$ cm$^3$s$^{-1}$.

Future dSphs analyses with the \textit{Fermi}-LAT telescope will benefit from larger statistics and potential discoveries of new ultra-faint dwarfs. At low energies the point spread function (PSF) sensitivity for the LAT instrument increases approximately as the square-root of the observation time, while at high energies, the PSF increases roughly linearly with time. The $\gamma$-ray bounds reported in Ref.~\cite{Ackermann:2015zua} used 6 years of \textsc{Pass8} \textit{Fermi} data taken from 15 dwarf spheroidals. Thus, we can conservatively estimate that with 15 years of \textit{Fermi} data and 3 times more dSphs discovered (45 dSphs) in the next few years, the LAT constraints will improve by a factor of $(\sqrt{15}/\sqrt{6})\times 3 \simeq 5$ compared to the current ones. 

As can be seen in Fig.~\ref{fig:sigmav}, the 15 years \textit{Fermi}-LAT forecast in the $W^{+}W^{-}$ channel indicates that future dSphs observations will be in significant tension with the set of favoured models found in Region I. Although the \textit{Fermi} collaboration have not yet released equivalent limits for $t\bar{t}$ final states, these should be comparable at the percentage level~\cite{Cirelli_cookbook} with those in the $b\bar{b}$ channel. We thus use the latest limits accordingly, and show that \textit{Fermi}-LAT dwarfs will also have the ability to test our $t\bar{t}$ solution (Region II). However, here an important remark is in order. As discussed in Ref.~\cite{Caloreetal:Taleoftails}, astrophysical uncertainties in the DM parameters can affect the expected $\gamma$-ray emission in a manner that makes the annihilation cross-section uncertain by a factor of $\sim 5$ up and down. Hence, both of our solutions could in principle still escape future \textit{Fermi}-LAT dwarfs limits if astrophysical uncertainties are taken into consideration. Also, as there is likely to be at least some millisecond pulsar contribution, the actual $\langle\sigma v\rangle$ could be correspondingly lower and so even harder to detect.     

Using the method presented in Ref.~\cite{Silverwood:2014yza}, we compute the 95\% confidence level upper limits on the annihilation cross section that will be achievable with the upcoming ground-based $\gamma$-ray observatory CTA~\cite{2011ExA....32..193A}, assuming annihilation into $W^+W^-$ and $t\bar{t}$ channels and the halo model described earlier in this paper. These limits use the 28 spatial bin morphological analysis, and include a systematic uncertainty of 1\% and the effects of the galactic diffuse emission. We find the 95\% confidence level upper limits by first calculating the best fit annihilation cross section, and then correctly increasing the cross section until $-2 \ln \mathcal{L}$ increases by 2.71 whilst profiling over the remaining signal model parameters. These limits are shown in Figure~\ref{fig:sigmav}, and show that observations towards the GC by CTA will be unable to confirm or exclude the SDFDM model as an explanation of the GCE. 

%
\begin{figure}[t]
\begin{center}
\includegraphics[scale=0.39]{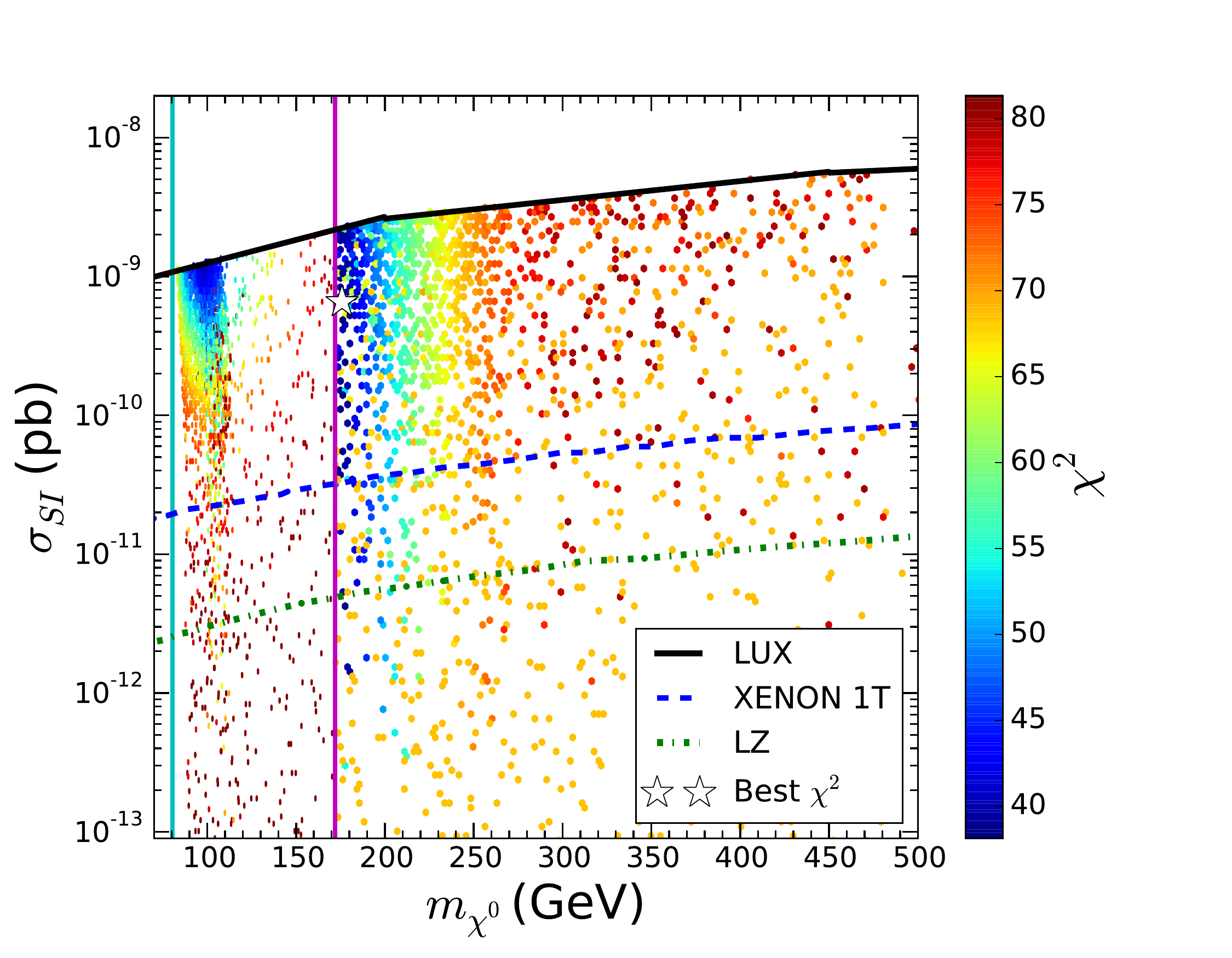}  
\includegraphics[scale=0.39]{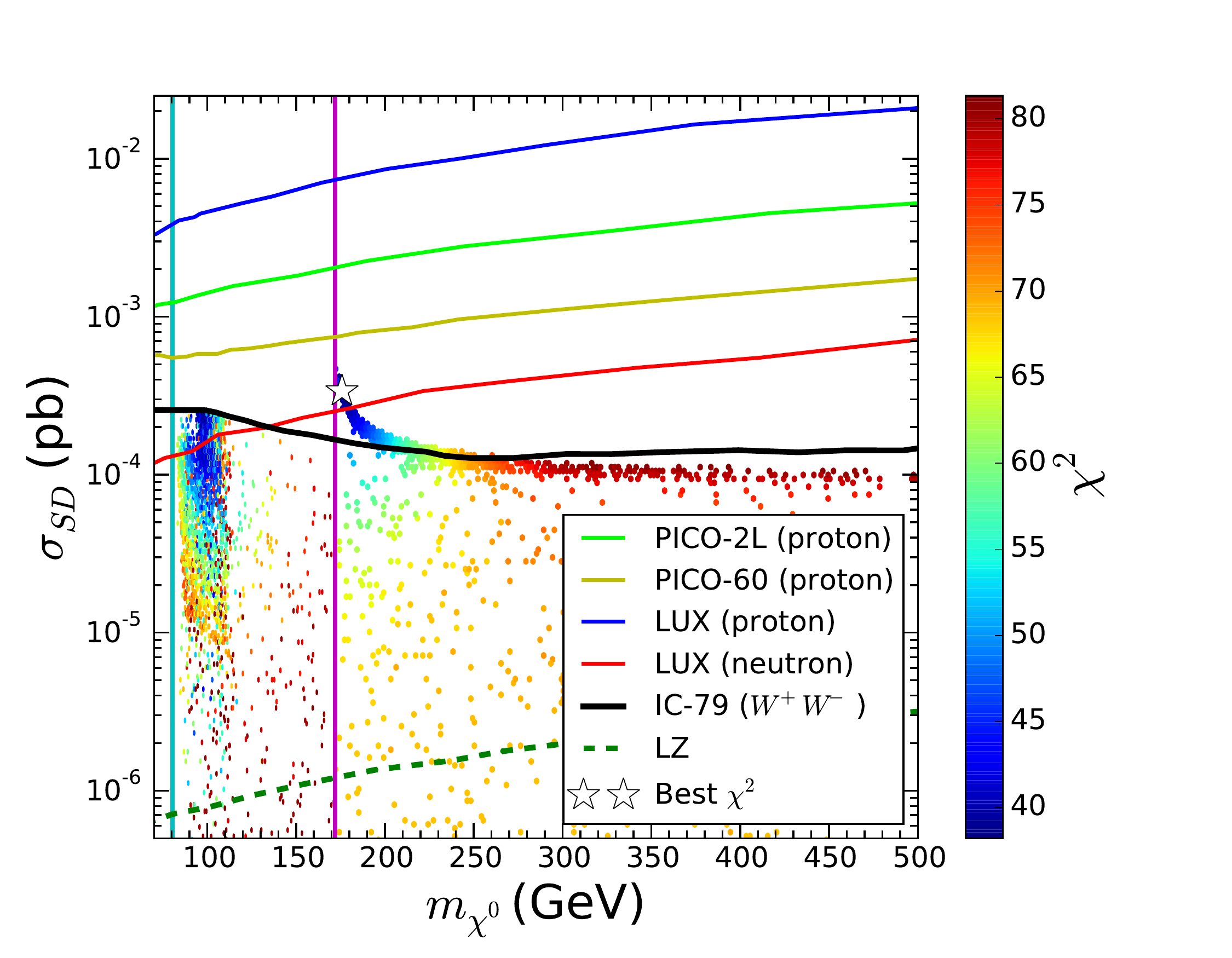} 
\caption{Spin-independent $\sigma_{SI}$ (left) and spin-dependent $\sigma_{SD}$ (right) direct detection cross sections in the SDFDM model in comparison to current and future direct detection limits. 
The left panel displays current limits from the LUX experiment (black solid line) and the expected limits from the forthcoming XENON-1T and LZ~\cite{Cushman:2013zza} experiments (blue dashed and green dot-dashed lines).
The right panel shows the IceCube limits in the $W^+W^-$ channel (black solid line) from null observations of the sun,  the PICO-2L~\cite{Amole:2016pye} (green light solid line) and PICO-60~\cite{Amole:2015pla} (yellow solid line) limits as well as the LZ sensitivity (green dashed line). The most recent constraints from LUX~\cite{Akerib:2016lao} (red and blue solid lines) are also overlaid. 
The star is the best-fitting model obtained from our scan. 
Vertical lines and color code are the same as in Fig.~\ref{fig:parameters}.
}
\label{fig:sigma-SI-SD}
\end{center}
\end{figure}
%
The SDFDM model can also be tested through direct dark matter detection searches. This results from either the spin-independent (SI) or spin-dependent (SD) scattering of the $\chi^0$ particle off a target nucleus. Fig.~\ref{fig:sigma-SI-SD} displays the predicted SI and SD cross sections for our model set together with several present and anticipated experimental constraints. Namely, we overlaid the upper limits from the LUX experiment, and the expected limits from XENON-1T and LZ~\cite{Cushman:2013zza}. As can be seen, these future experiments, in particular LZ, will be able to cut deeply into the model set and confirm or rule out the DM explanation of the GCE if it is the only extended source emitting high energy photons in the GC. We also note that available constrains from IceCube are just on the edge of probing the set of models that could account for the excess. In fact, the most recent limits on the spin-dependent WIMP-nucleon elastic cross-section from LUX~\cite{Akerib:2016lao} have begun to disfavor the best fit region. This is per se, a great example of the importance of a combined effort of different search techniques in the quest for dark matter.

\section{Conclusions}
\label{sec:conclusions}

In this work we have entertained the possibility of finding model points in the SDFDM model that can explain the GCE while being in agreement with a multitude of different direct and indirect DM detection constrains. We found two viable regions: (i) DM particles present in the model with masses of $\sim 99$ GeV annihilating mainly into $W$ bosons with branching ratios greater than $\sim 70\%$, (ii) and a second region where the DM particle mass is in the range $\sim (173-190)$ GeV annihilating predominantly into the $t\bar{t}$ channel with branching ratios greater than $\sim 90\%$. Our analysis assumed that the DM is made entirely out of the lightest stable particle $\chi^0$ of the SDFDM model. Despite this being a very restrictive assumption, we have demonstrated that there exist models capable of accounting for the GeV excess in the GC that can be fully tested by the forthcoming XENON-1T and LZ experiments as well as by future \textit{Fermi}-LAT observations in dwarf galaxies. Interestingly, the most recent limits presented by LUX are able to probe a fraction of the good fitting models to the GCE found in this work. We also showed through realistic calculations of CTA performance when observing the GC that this instrument will not have the ability to confirm the SDFDM model if it is causing the GCE.

\section{Acknowledgments}
We would like to thank Chris Gordon, Marta L. Sánchez and Christoph Weniger for useful discussions. D.R. and O.Z. have been partially supported by UdeA through the Grants Sostenibilidad-GFIF, CODI-2014-361 and CODI-IN650CE, and COLCIENCIAS
through the Grants No. 111-556-934918 and 111-565-842691. A.R. is supported by COLCIENCIAS (Doctorado Nacional-6172) and acknowledges the hospitality of Universit\'e Libre de Bruxelles while this work was being completed.

\bibliographystyle{jhep}
\bibliography{susy}

\end{document}